\documentclass{article}
\usepackage{amsmath}
\usepackage{graphicx}
\usepackage{geometry}

\def\pr{\mathrm{Pr}}
\def\A{\mathcal{A}}
\def\S{\mathcal{S}}
\def\I{\mathcal{I}}
\def\R{\mathcal{R}}
\def\J{\mathcal{J}}

\begin{document}

\author{Erik Volz \footnote{Department of Integrative Biology, University of Texas-Austin, 1 University Station, C0930, Austin, TX 78712. To whom correspondence should be addressed. E-mail: erik.volz@mail.utexas.edu} \and Lauren Ancel Meyers \footnote{Department of Integrative Biology, University of Texas-Austin, 1 University Station, C0930, Austin, TX 78712.} }


\title{SIR epidemics in dynamic contact networks}
\date{Received: March 14, 2007}

\maketitle

\begin{abstract}
Contact patterns in populations fundamentally influence the spread of infectious diseases. Current mathematical methods for epidemiological forecasting on networks largely assume that contacts between individuals are fixed, at least for the duration of an outbreak. In reality, contact patterns may be quite fluid, with individuals frequently making and breaking social or sexual relationships. Here we develop a mathematical approach to predicting disease transmission on dynamic networks in which each individual has a characteristic behavior (typical contact number), but the identities of their contacts change in time. We show that dynamic contact patterns shape epidemiological dynamics in ways that cannot be adequately captured in static network models or mass-action models. Our new model interpolates smoothly between static network models and mass-action models using a mixing parameter, thereby providing a bridge between disparate classes of epidemiological models. Using epidemiological and sexual contact data from an Atlanta high school, we then demonstrate the utility of this method for forecasting and controlling sexually transmitted disease outbreaks. \\
{\footnotesize Keywords:Infectious disease | SIR | Networks | Syphilis}
\end{abstract}

\section{Introduction}

	Most epidemic models incorporate a homogeneous mixing assumption, sometimes called the law of mass action~\cite{ross1910pm,and,die}, whereby the rate of increase in epidemic incidence is proportional to the product of the number of infectious and the number of susceptible individuals. This assumption has been relaxed in some compartmental~\cite{vandendriessche2002rna} and meta-population models~\cite{lloyd1996she,finkenstadt1998edm,grenfell2001twa,watts2005mre}, but not eliminated. 
	The mass action assumption is robust in the sense that it is consistent with several scenarios for the individual-to-individual transmission of disease. In particular, it is equivalent to a model in which all individuals in a population make contact at an identical rate and have identical probabilities of disease transmission to those contacts per unit time. 
	Although this assumption is unrealistic, it facilitates mathematical analysis and, in some cases, offers a reasonable approximation. 
	 
	Populations can be quite heterogeneous with respect susceptibility, infectiousness, contact rates, or number of partners, and simple homogeneous mixing models do not allow for extreme variation in host parameters. New network-based mathematical methods capture some, but not all, aspects of population heterogeneity ~\cite{callaway2000nra,strog1,newm2,lilj1,an,andeMay2,newm1}. Ideally, an epidemic model would incorporate the following realities of human-to-human contacts:
\begin{itemize}
	\item  A given individual has contact with only a finite number of other individuals in the population at any one time, and contacts which can result in disease-transmission are usually short and repeated events.
	\item The number and frequency of contacts between individuals can be very heterogeneous. 
	\item The numbers and identities of an individual's contacts will change as time goes by. 
\end{itemize}
The first two points have been addressed previously by static network models~\cite{newm1,eameKeel1,meyers2005nta,meyers2006ped}. We will focus on the third point, and introduce a modeling framework that allows an individual's contacts to change in time. 

Concurrent and serial contacts were first shown to be important to HIV transmission dynamics in~\cite{watts1992icp,dietz1988tdh}, and have been modeled using high-dimensional pair-approximation methods~\cite{altmann1995sir,altmann1998dli, eames2004mna} and moment closure methods on dynamic contact networks~\cite{bauch2002voa}. Here, we introduce a low-dimensional system of non-linear ODEs to model susceptible-infected-recovered (SIR) epidemics in a simple class of dynamic networks. We use this model to characterize the impacts of population heterogeneity and contact rates on epidemic dynamics. We show further that the model reproduces basic classes of epidemiological models, such as the standard mass-action SIR model and static network model, as the parameter which controls population mixing varies.

\section{Neighbor-Exchange model}
	Human contact patterns are potentially complex, as the numbers and intensity of contacts can vary considerably across a population. Furthermore, contacts are transitory events such that the identities of one's contacts change in time. To capture such heterogeneity, we introduce the \emph{neighbor-exchange} (NE) model as a simple extension of a static contact network model. In this model, an individual's number of concurrent contacts remains fixed while the composition of those contacts changes at a specified rate. The model assumes that at any given time, an individual will be {in contact} with an individual-specific number of neighbors with whom disease transmission is possible. Each contact is temporary, lasting a variable amount of time before coming to end, at which point the neighbor is replaced by a different individual.
	
Let the population of interest consist of $n$ individuals, each of which falls into one of three exclusive states: susceptible, infectious, or recovered. At some time $t$, an individual $ego$ will have $k_{ego}$ contacts with other individuals (i.e., \emph{alters}): $(ego, alter_1), (ego, alter_2), \cdots, (ego,alter_k)$. Only undirected contact networks will be considered such that if there exists a  contact $(ego,alter)$ there will also be a contact $(alter, ego)$. 
In network terminology a directed link, denoted $(ego, alter)$, is called an \emph{arc}. An undirected link, denoted $\{ego,alter\}$, is called an \emph{edge}. The \emph{degree} of a node $ego$ is the number of edges connected to the node. The term \emph{contact} will specifically be used to denote a directed arc in the network, where two arcs correspond to each undirected edge. The \emph{k-degree} of a node will be the number of concurrent contacts to/from the node.
	
	The neighbor exchange model assumes that the identities of a node's neighbors will continually change while the total number of current neighbors remains constant. This occurs through an exchange mechanism in which the destination nodes of two edges are swapped. For example, two nodes $ego$ and $ego'$ with distinct contacts $(ego,alter)$ and $(ego',alter')$ may exchange contacts such that these are replaced with $(ego,alter')$ and $(ego',alter)$. There are two edges and four contacts involved in each edge-swap. The fate of each edge and contact is summarized in the following pseudo-chemical equation:
		\begin{equation}
			\label{eqn:edgeswap}
			\begin{split}
			\{ego,alter\} + \{ego', alter'\} \longrightarrow \{ego,alter'\} + \{ego', alter\} \\
			(ego, alter) + (ego', alter') \longrightarrow  (ego,alter') + (ego', alter) \\
			(alter, ego) + (alter', ego') \longrightarrow  (alter,ego') + (alter', ego) \\
			\end{split}
		\end{equation}
	In the model, any given contact $(ego,alter)$ will be reassigned to $(ego,alter')$ at a constant rate $\rho$. Equivalently, edges are swapped at a rate $2\rho$.

	\begin{table}
		\begin{center}
		\begin{footnotesize}
		\caption{Notation for epidemic and network parameters.}
		\begin{tabular}{ll}
		\hline
			$S$ & The fraction of population susceptible. \\
			$I$ & The fraction of population infectious. \\
			$R$ & The fraction of population recovered. \\
				$\S$ & Set of susceptible nodes.\\
				$\I$ & Set of infectious nodes. \\
				$\R$ & Set of recovered nodes.\\
			$J$ & Cumulative epidemic incidence ($J = I+R$).  \\
			$r$ & Transmission rate. \\
			$\mu$ & Recovery rate. \\
			$\rho$ & Mixing rate. \\
			$k_{ego}$  & Degree of node $ego$. \\
			$p_k$ & The fraction of nodes with degree $k$. \\
		\hline
		\end{tabular}
		\end{footnotesize}
		\end{center}
	\end{table}	
	
	For mathematical tractability, we make a few simplifying assumptions about the epidemic process. First, for the duration of a contact, infectious individuals transmit disease to neighbors at a constant rate $r$. Second, infectious individuals become recovered at a constant rate $\mu$.
	
	We also simplify the mathematics by considering only the simplest category of heterogeneous networks: semi-random networks which are random with respect to a given degree distribution~\cite{mollReed1,moRe95}. The degree distribution will have density $p_k$, which is the probability that a node chosen uniformly at random has $k$ concurrent contacts. Note that a node's degree never changes during an edge-swap and thus the degree distribution is preserved.
	
		The simultaneous epidemic and network dynamics described above collectively determine the neighbor-exchange model. An even more realistic model would allow the number $k$ of concurrent contacts of a node to vary stochastically, but the current model offers a valuable first step toward understanding epidemiological processes on dynamic host networks. 
	
	\subsection{Dynamics}
		We will expand on the dynamic probability generating function (PGF) techniques first introduced in~\cite{volzJMB} to model SIR-type epidemics in static networks. These techniques are powerful and are easily extended to consider dynamic contact networks. We start with an overview of the basic model and then introduce additional terms that model the neighbor exchange process.  
		
		The concurrent degree distribution $p_k$ will be generated by the PGF
		\begin{equation}
			\label{eqn:g}
			g(x) = p_0 + p_1 x + p_2 x^2 + p_3 x^3 + \cdots
		\end{equation}
		The dummy variable $x$ in this equation serves as a placeholder for dynamic variables.
		
		Let $\A_X$ be the sets of contacts (arcs) where $ego$ is in the set $X$. We will consider the sets of all susceptible, infected and recovered nodes, denoted $X=\S$, $X=\J$, and $X=\R$, respectively. $M_X=\#\{\A_X\}/\#\{\A\}$ is then defined as the fraction of total contacts in set $\A_X$. Now let $\A_{XY}$ be the set of contacts such that $ego$ is the set $X$ and $alter$ is in the set $Y$, and $M_{XY} = \#\{\A_{XY}\}/\#\{\A\}$ be the fraction of total contacts in set $\A_{XY}$. For example, $M_{\S\S}$ is the fraction of all arcs in the network that connect two currently susceptible nodes.  
			\begin{table}
				\begin{footnotesize}
				\caption{Notation for network and epidemic quantities. \label{tab:def2}}
				\begin{tabular}{ll}
				\hline
				$\A_X$  & Set of contacts $(ego,alter)$ s.t. $ego\in X$. \\
				$\A_{XY}$  &  Set of contacts $(ego,alter)$ s.t. $ego\in X$  and $alter\in Y$. \\
				$M_X$  &  Fraction of contacts in set $\A_X$. \\
				$M_{XY}$ & Fraction of contacts in set $\A_{XY}$.\\
				$p_I = M_{\S\I}/M_\S$ & Fraction of contacts from susceptibles which go to infectious nodes.\\
				$p_S = M_{\S\S}/M_\S$ & Fraction of contacts from susceptibles which go to other susceptible nodes.\\
				\hline
				\end{tabular}
				\end{footnotesize}
			\end{table}
		
		The following derivation assumes that each contact of a susceptible node $(ego,alter)\in \A_{\S}$ has a uniform probability that $alter\in \I$, denoted $p_I=M_{\S\I}/M_{\S}$, and a uniform probability $p_S=M_{\S\S}/M_\S$ that $alter\in \S$. A degree $k$ susceptible node has an expected number $k p_I$ contacts with infectious nodes; and, in a small time $dt$, an expected number $r k p_I ~dt$ of a degree $k$ susceptible nodes' contacts will transmit disease to that node. The instantaneous hazard of infection for a degree $k$ susceptible node is then given by
		\begin{equation}
			\label{eqn:lambdak}
			\lambda_k(t) = r k p_I(t)
		\end{equation}
		
		Let $u_k(t)$ denote the fraction of degree $k$ nodes remaining susceptible at time $t$. Equation ~\ref{eqn:lambdak} implies
		\begin{equation}
			\label{eqn:uk}
			\begin{split}
			{\displaystyle u_k(t) = \exp \{ \int_{\tau=0}^{t} -\lambda_k(\tau) d\tau  \}   }\\
			{\displaystyle = \exp \{ \int_{\tau=0}^{t} -r k p_I(t) d\tau  \}   } \\
			{\displaystyle = \exp \{ \int_{\tau=0}^{t} -r p_I(t) d\tau  \}^k   }
			\end{split}
		\end{equation}
		
		Now let $\theta=u_1(t)$ be the fraction of degree $k=1$ nodes in the network which remain susceptible at time $t$. Equation~\ref{eqn:uk} implies that $u_k = \theta^k$. (Henceforth, variables that are clearly dynamic, like $\theta$, appear without a time $(t)$ variable.)  
		
		We use the PGF for the network degree distribution (equation~\ref{eqn:g}) to calculate the fraction of nodes that remain susceptible at time $t$. 
		\begin{equation}
			\label{eqn:S}
		 S = p_0 + u_1 p_1 + u_2 p_2 +\cdots  = \sum_k p_k \theta^k = g(\theta).
		\end{equation}
		
		The dynamics of $\theta$ in equation~\ref{eqn:S} are given by
		\begin{equation}
			\label{eqn:theta}
			{\displaystyle \dot{\theta} = \frac{d}{dt} u_1 = -\lambda_1 \theta  = - r p_I \theta   }.
		\end{equation}
		
		Unfortunately, this does not form a closed system of differential equations, as equation~\ref{eqn:theta} depends on the dynamic variable $p_I$. From the definition of $p_I$ we have
		\begin{equation}
			\label{eqn:pidot1}
			\dot{p}_I = \frac{d}{dt} \frac{M_{\S\I}}{M_S} = \frac{\dot{M}_{\S\I}}{M_{\S}} - \frac{M_{\S\I} \dot{M}_S}{M_{\S}^2}. 
		\end{equation}
		
		To obtain the derivatives of $M_{\S\I}$ and $M_\S$, we observe that, in time $dt$, $-\dot{S}$ nodes become infectious, resulting in modifications to the sets $\A_{\S\I},\A_{\I\S},$ and $\A_\S$. In particular, any given arc from newly infected individual was formerly in one of $\A_{\S\S}$, $\A_{\S\I}$, or $\A_{\S\R}$, and is now in one of $\A_{\I\S}$, $\A_{\I\I}$, or $\A_{\I\R}$. The $-\dot{S}$ new infecteds are not selected uniformly at random from the susceptible population, but rather with probability proportional to the number of contacts to infectious nodes. 

We pause for a few definitions. First, it is useful to break the degree of a node into three quantities: the number of contacts to currently susceptible, infected, and recovered nodes. We refer to these as \emph{X-degrees}, where $X=S, I$, or $R$, respectively. Second, imagine following a randomly chosen contact to its alter node and counting all the edges emanating from that node, except for the one on which we arrived. We call the resulting total quantity the \emph{excess degree} of the node, and the resulting neighbor-specific quantities \emph{excess X-degrees} of the node, where $X$ indicates one of the three possible disease states. 
	
We introduce the notation $\delta_{X,Y}(Z)$ to represent the average excess $Z$-degree of nodes currently in disease state $X$ selected by following a randomly chosen arc from the set $\A_{YX}$-- in other words, the excess $Z$-degree of a node of type $X$ selected with probability proportional to the number of contacts to nodes of type $Y$. We further define $\delta_{X,Y}$ to be the average (total) excess degree of nodes currently in disease state $X$ selected by following a randomly chosen arc from the set $\A_{YX}$. 
For example, imagine first randomly choosing an arc from $\A_{\I\S}$, then following that arc to its destination (susceptible) node, and finally counting all of the other edges emanating from that node (ignoring the one along which we arrived). Then $\delta_{S,I}(S)$, $\delta_{S,I}(I)$, and $\delta_{S,I}(R)$ give the average number of contacts to other susceptible, infected and recovered nodes chosen in this way, respectively; and $\delta_{S,I}$ gives the average total number of contacts emanating from nodes chosen in this way.
	
	Using this notation, the equations for $\dot{M}_{\S\I}$ and $\dot{M}_\S$ are as follows (for more details, see~\cite{volzJMB}).
		\begin{eqnarray}
			\dot{S} = \frac{d}{dt} g(\theta) = \dot{\theta} g'(\theta) = -r p_I \theta g'(\theta) \\
			\dot{M}_{\S\I} = ( (-\dot{S}) \delta_{S,I}(S) - (-\dot{S}) \delta_{S,I}(I) )/g'(1) - (r+\mu) M_{\S\I} \\
			\dot{M}_{\S\S} = -2 (-\dot{S}) \delta_{S,I}(S) / g'(1) \\
			\dot{M}_{S} = \frac{d}{dt} \theta g'(\theta) / g'(1) = -(r p_I \theta g'(\theta) + r p_I \theta^2 g''(\theta))/g'(1)
		\end{eqnarray}
		The calculations of the $\delta_{X,Y}(Z)$ are straightforward, and based on the current degree distribution of susceptible nodes. The calculations are given in the supplement~\cite{supp} and in~\cite{volzJMB}, and result in the following.
		\begin{eqnarray}
			\delta_{S,I}(I) = p_I \theta g''(\theta)/g'(\theta) \\
			\delta_{S,I}(S) = p_S \theta g''(\theta)/g'(\theta) 
		\end{eqnarray}
		
		Combining the equations for $M_\S,\dot{M}_\S,$ and $\dot{M}_{\S\I}$ yields the dynamics of $p_I$ in terms of the parameters $r$ and $\mu$, the PGF $g(\cdot)$, and the dynamic variable $p_S$. The resulting model is given in table~\ref{tab:netsir1}. The dynamics for $p_S$ complete the model, and can be derived analogously to the equation for $p_I$ (see the supplement~\cite{supp} for details). 		
		
		\begin{table}
			\begin{center}
			\begin{footnotesize}
			\caption{System of equations used to model the spread of an SIR epidemic in a static semi-random network. \label{tab:netsir1}}
			\begin{tabular}{ll}
			\hline
				$\dot{\theta}=$ & $-r p_I \theta$ \\
				$\dot{p}_S=$ & $r p_S p_I \left( 1-\theta g''(\theta)/g'(\theta) \right)$ \\
				$\dot{p}_I=$ & $r p_I p_S \theta g''(\theta)/g'(\theta) - p_I (1-p_I) r - p_I \mu$ \\
			\hline
				$S=$ & $g(\theta)$ \\
			\hline
			\end{tabular}
			\end{footnotesize}
			\end{center}
		\end{table}

	\subsubsection{Dynamic contact networks}
		We now extend the model given in table~\ref{tab:netsir1} to allow neighbor exchanges (NE) at a rate $\rho$. 
		First consider $\theta$. An edge-swap (equation~\ref{eqn:edgeswap}) will affect the arrangement of edges among susceptible, infectious, or recovered nodes, however it will never directly cause a node to change its disease state. The dynamics of $\theta$ are only indirectly influenced by NE dynamics, through changes to $p_I$, and thus equation ~\ref{eqn:theta} still holds. 
		
		NE does, however, affect the composition of contacts between susceptible and infectious nodes. We postulate that the equations for $\dot{p}_I$ and $\dot{p}_S$ can be expressed in the following modified forms.
		\begin{eqnarray}
			\label{eqn:dotp2}
		 \dot{p}_I = r p_I p_S \theta g''(\theta)/g'(\theta) - p_I (1-p_I) r - p_I \mu + f_I(p_I,M_I)\\
		 \dot{p}_S = r p_S p_I \left( 1-\theta \frac{g''(\theta)}{g'(\theta)} \right) + f_S(p_S, M_S) 
		\end{eqnarray}
		where the functions $f_x(\cdot,\cdot)$ represent the contribution of NE dynamics to the system. 
		
		NE dynamics can both decrease or increase the values of $p_I$ and $p_S$. First consider the decrease of $p_I$ due to NE dynamics.  
		\begin{itemize}
			\item At rate $\rho$, a given contact $(ego,alter)$ will transform to $(ego, alter')$.
			\item Given that $(ego,alter)\in \A_S$, $alter\in \I$ with probability $p_I$.
			\item With probability $1-M_I$, $alter' \notin \I$.
		\end{itemize}
		Thus $p_I$ will be decreased by NE dynamics at a rate $\rho p_I (1-M_I)$.
		
		Similarly, $p_I$ can increase due to NE as follows. 
		\begin{itemize}
			\item At rate $\rho$, a given contact $(ego,alter)$ will transform to $(ego, alter')$.
			\item Given that $(ego,alter)\in \A_S$, $alter\notin \I$ with probability $1-p_I$.
			\item With probability $M_I$, $alter' \in J$.		
		\end{itemize}
		Thus $p_I$ is increased by NE dynamics at rate $\rho (1-p_I) M_I$. We add the positive and negative contributions to calculate the total influence of NE on $p_I$.
		\begin{equation}
			\label{eqn:fi}
			f_I(p_I,M_I) = \rho \left( M_I - p_I \right)
		\end{equation}
		
		By similar reasoning, we determine $f_S(p_S, M_I)$.
		\begin{equation}
			\label{eqn:fs}
			\begin{split}
			f_S(p_S, M_S) = \rho \left( \pr[alter\notin \S]\times \pr[alter'\in S] - \pr[alter\in \S]\times \pr[alter'\notin S]\right) \\
			= \rho\left( M_S - p_S \right) = \rho \left( g'(\theta)/g'(1) - p_S \right)
			\end{split}
		\end{equation}
		
		The complete system of NE-adjusted equations is reported in table~\ref{tab:netsir2}. The dynamic variable $M_I$, however, appears in equation~\ref{eqn:fs} and cannot be put in terms of the variables $\theta, p_S,$ and $p_I$. Therefore a fourth dynamical equation must be included for $M_I$, which is listed in table~\ref{tab:netsir2}. Fortunately, the dynamics are very straightforward. Recall that $\delta_{S,I}$ denotes the average excess degree of susceptibles chosen by following randomly selected arcs between infectious and susceptible nodes, and thus indicates the average excess degree for individuals who become infected in a small time interval $dt$. We add one ($\delta_{S,I}+1$) to obtain the average degree for such newly infected nodes. Then, recalling that $M_I$ decays at rate $\mu$, we have 
		\begin{equation}
			\label{eqn:dotmi}
			\dot{M}_I = (-\dot{S}) (\delta_{S,I} + 1) - \mu M_I
		\end{equation}
		
		\begin{table}
			\begin{center} 
			\begin{footnotesize}
			\caption{System of equations used to model the spread of an SIR type epidemic in a dynamic semi-random network with stochastic exchange of neighbors at constant rate $\rho$. \label{tab:netsir2} }
			\begin{tabular}{ll}
			\hline
				$\dot{\theta}=$  &  $- r p_I \theta$ \\	
				$ \dot{p_S}=$ & ${\displaystyle r p_S p_I \left( 1-\theta g''(\theta)/g'(\theta) \right) + \rho \left( g'(\theta)/g'(1) - p_S \right) }$ \\ 
				$ \dot{p_I}=$  &  ${\displaystyle r p_I p_S \theta g''(\theta)/g'(\theta) - p_I (1-p_I) r - p_I \mu + \rho \left( M_I - p_I \right) }$ \\				
				$\dot{M_I}=$ & $ {\displaystyle -\mu M_I + r p_I \left( \theta^2 g''(\theta) + \theta g'(\theta) \right) /g'(1) }$\\
			\hline
				$S =$ &  $g(\theta)$ \\
			\hline
			\end{tabular}
			\end{footnotesize}
			\end{center}
		\end{table}
		
	All of the results that follow will assume that a very small fraction  $\epsilon$ of nodes are initially infected, and thus there is only a very small probability that two initially infected individuals contact each other.  Then we anticipate the following initial conditions~\cite{volzJMB}:
	\begin{eqnarray}
		\label{eqn:ic}
			\theta = M_I = \epsilon \\
			p_I = \epsilon /  (1-\epsilon) \\
			p_S = (1-2\epsilon) / (1-\epsilon)
	\end{eqnarray}
		
\subsection{Convergence to a mass action model\label{sec:mamodel}}
	In the limit of large mixing rate ($\rho\rightarrow \infty$), the probability of being connected to a susceptible, infectious, or recovered node is directly proportional to the number of edges emanating from nodes in each state. Referring to table~\ref{tab:netsir2}, it is clear that $p_I$ converges instantly to $M_I$ and $p_S$ converges instantly to $M_S$. Here, we show that, as the mixing rate grows, the underlying network structure becomes irrelevant and the model converges to a mass action model.
	
	To see this, we replace every occurrence of the variable $p_I$ in the system of equations in table~\ref{tab:netsir2} with $M_I$. Then
	\begin{equation}
		\label{eqn:thetadotrm}
		\dot{\theta} = -r M_I \theta, 
	\end{equation}
	and, 
	\begin{equation}
		\label{eqn:midotrm}
		{\displaystyle \dot{M}_I = \frac{r M_I}{g'(1)} \left(\theta g'(\theta) + \theta^2 g''(\theta) \right) - \mu M_I   }
	\end{equation}
	Neither equations~\ref{eqn:thetadotrm} or~\ref{eqn:midotrm} depend on $p_S$, and thus together form a closed system of equations which describe the epidemic dynamics. These equations incorporate arbitrary heterogeneity in contact rates, but no longer consider an explicit contact network. When we assume that contact rates are homogeneous throughout the population, than these equations are equivalent to a simple SIR compartmental model. To illustrate, we retrieve the standard SIR dynamics by setting $g(x)=x$, which means that every node has exactly one concurrent contact. In such a population, the number of arcs to infectious individuals is exactly equal to the number of infectious nodes, that is, $M_I=I$. Then, substituting into equations~\ref{eqn:thetadotrm} and~\ref{eqn:midotrm}, we reproduce the standard equations:
	\begin{eqnarray}
		\label{eqn:rmeqn2}
		S = g(\theta) = \theta \\
		\dot{\theta} = -r I \theta = -r I S \\
		\dot{M}_I = r M_I \theta - \mu M_I = r I S - \mu I
	\end{eqnarray}
	
	Equations~\ref{eqn:thetadotrm} and~\ref{eqn:midotrm} are potentially extremely useful, as they incorporate arbitrary heterogeneity in a system of equations no more complex than the standard compartmental SIR model. 
	
\begin{figure}
			\begin{center}
				\includegraphics[width=.65\textwidth]{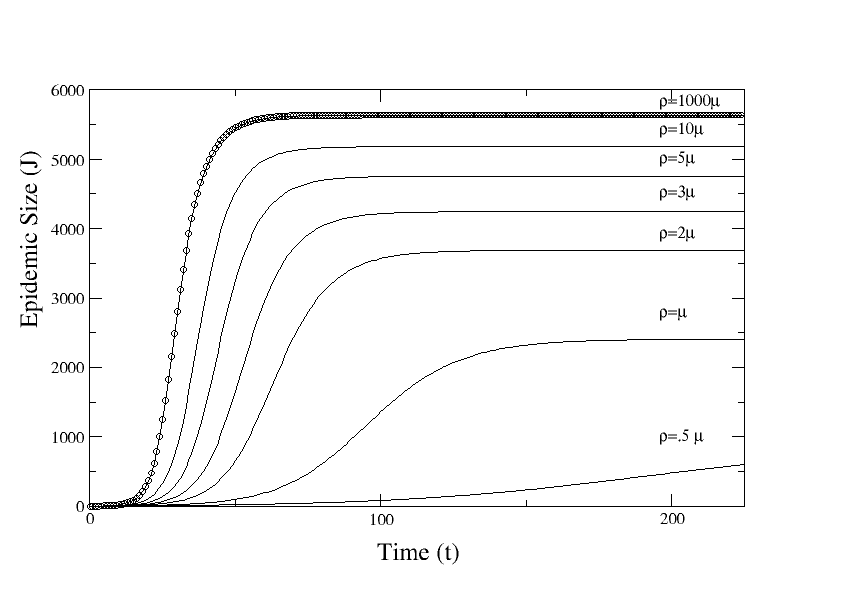}
				\caption{ Trajectories of the NE model (solid lines, table~\ref{tab:netsir2}) are compared for several values of the mixing rate ($\rho$) to an analogous mass-action model (circles, equations~\ref{eqn:thetadotrm} and~\ref{eqn:midotrm}). The degree distribution is Poisson ($z=1.5$) and $r=\mu=0.2$. \label{fig:rmConvergePoisson}}
			\end{center}
		\end{figure}

	Figure~\ref{fig:rmConvergePoisson} demonstrates the convergence of the NE model to the corresponding mass-action model for a Poisson degree distribution ($r=\mu= 0.2, z=1.5$). The circles indicate the solution to the mass action model (equations~\ref{eqn:thetadotrm} and~\ref{eqn:midotrm}). We observe that the convergence is quite rapid as $\rho$ is increased in multiples of $\mu$. This supports the common assumption that the mass-action model is a reasonable approximation for populations marked by many short-duration contacts. 
		
\section{Stochastic simulations\label{sec:sim}}
		To test the NE model, we compare its predictions to stochastic simulations of an analogous epidemic process in networks.
		We first generate semi-random networks using the configuration model~\cite{moRe95}. The epidemic simulations then proceed as follows:
		\begin{enumerate}
			\item One node is selected uniformly at random from the population to be \emph{patient zero}, the first infected individual.
			\item Each contact $a\in \A$ has an exchange time $\Delta t_E(a)$ drawn from an exponential distribution (parameter $\rho$). This time is added to a queue.
			\item When a node $v$ is infected at time $t$, a time of infection $\Delta t_I(v,a)$ is drawn from an exponential distribution (parameter $r$) for each contact $a=(ego,alter)$ such that $v$ is $ego$. The time $t+\Delta t_I(v,a)$ is added to a queue.
			\item When a node $v$ is infected at time $t$, a time of recovery $\Delta t_R(v)$ is drawn from an exponential distribution (parameter $\mu$) and is assigned to $v$. The time $t+\Delta t_R(v)$ is added to a queue.
			\item When $t' = t+\Delta t_E(a)$ is the earliest time in the queue, an edge-swap is performed, as per equation~\ref{eqn:edgeswap}. The first edge involved in the swap corresponds to the contact $a$. The second edge is selected by choosing a unique element out of the of all edges uniformly at random.  Then a new time $t' + \Delta t_E(a)$ is drawn and added to the queue. 
			\item When $t' = t + \Delta t_I(v,a)$ is the earliest time in the queue, a transmission event will occur, providing $v$ has not recovered. Node $v$ transmits to whatever node is occupying the position of $alter$ at that time, causing $alter$ to change state to $\I$ if currently susceptible. If a transmission event occurs, a new time $t'+\Delta t_I(v,a)$ is drawn and added to the queue. 
			\item When $t+\Delta t_R(v)$ is the earliest time in the queue, the corresponding node $v$ enters a recovered state such that any transmission event with $v=ego$ is removed from the queue. 
		\end{enumerate}
		This process continues until there are no more transmission events in the queue. 
		
\begin{figure}
			\begin{center}
				\includegraphics[width=.65\textwidth]{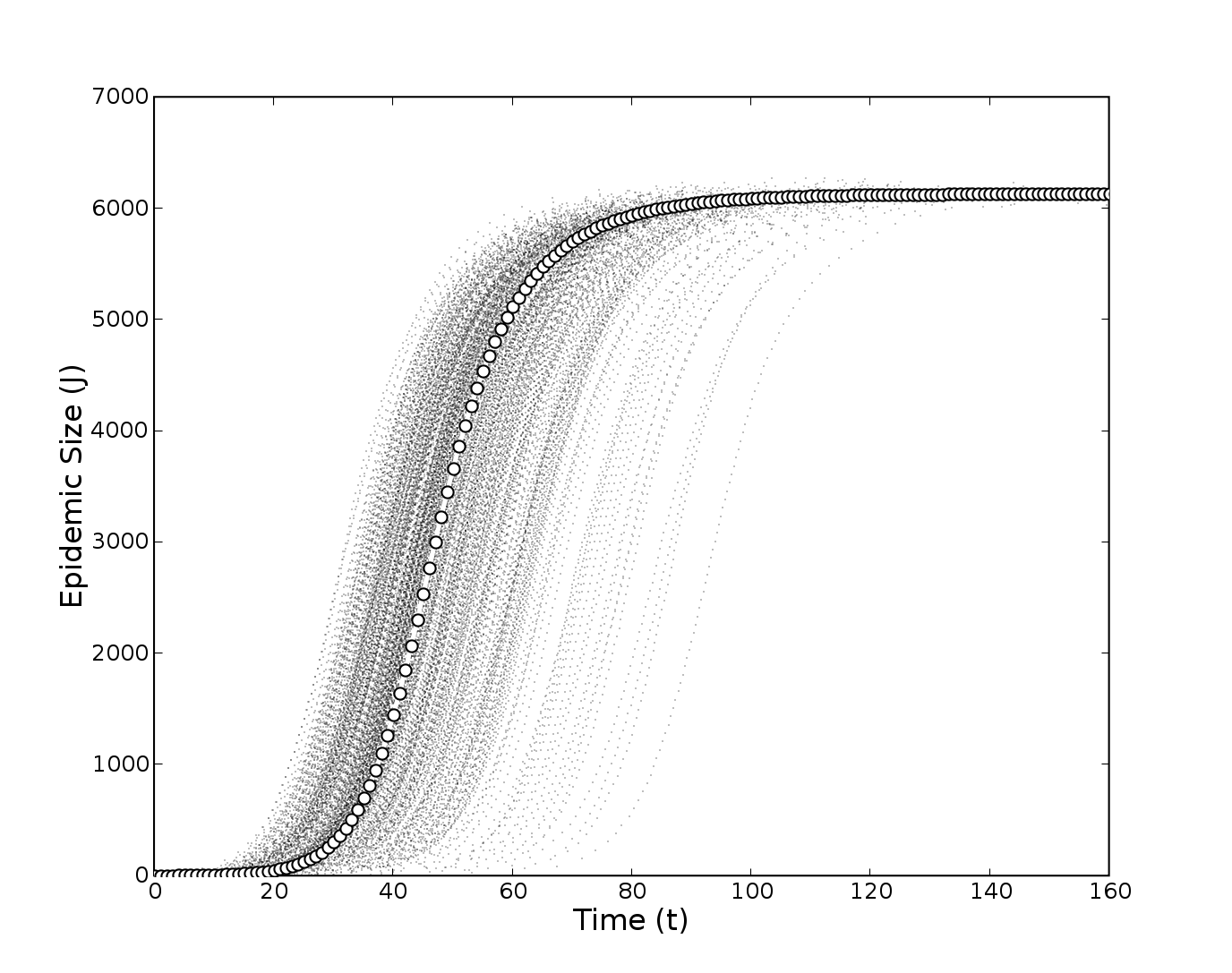}
				\caption{1000 stochastic simulations (small dots) are compared to the predicted trajectory of a NE epidemic (large dots) in a Poisson network ($z=1.5$).$r=0.2,\mu=0.1$, and $\rho=0.25$.\label{fig:poisSimFig}}
			\end{center}
		\end{figure}

\begin{figure}
			\begin{center}
				\includegraphics[width=.65\textwidth]{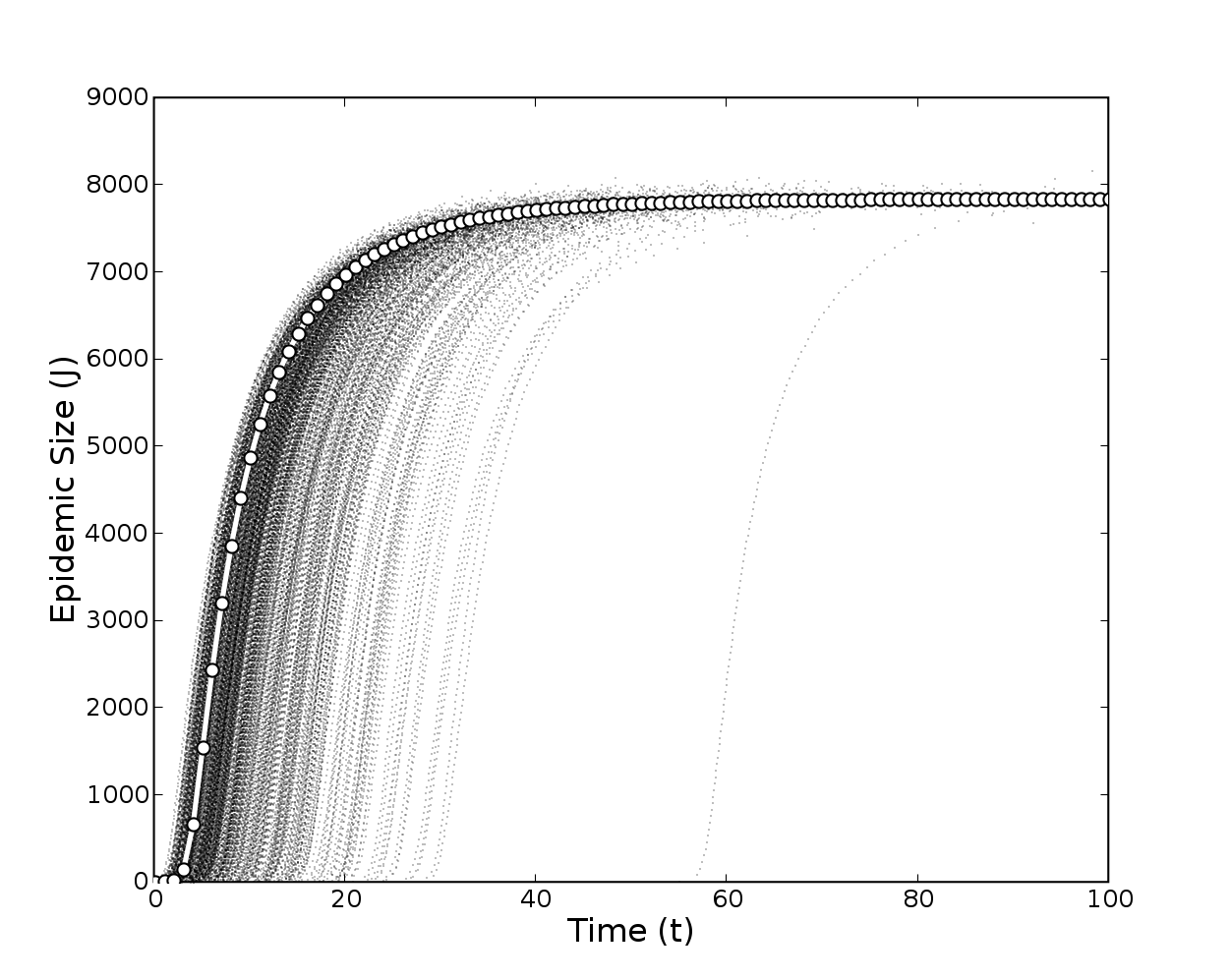}
				\caption{1000 stochastic simulations (small dots) are compared to the predicted trajectory of a NE epidemic (large dots) in a power-law network ($\alpha=2.1,\kappa=75$). $r=0.2,\mu=0.1$, and $\rho=0.20$.\label{fig:plSimFig}}
			\end{center}
		\end{figure}

		Figures~\ref{fig:poisSimFig} and~\ref{fig:plSimFig} show a comparison of one thousand stochastic simulations to the solution of the NE-SIR equations for two concurrent degree distributions:
		\begin{itemize}
			\item Poisson. $p_k = z^k e^{-z} / k!$, $k\ge 0$. This is generated by $g(x) = e^{z(x-1)}$.
			\item Power law with cutoff. $p_k = k^{-\alpha} / \sum_{i=1}^{\kappa} i^{-\alpha}$, $k\ge 1, k\le \kappa$. This is generated by $g(x) = \sum_k k^{-\alpha} x^k / \sum_{i=1}^{\kappa} i^{-\alpha}$. 
		\end{itemize}
		Figure~\ref{fig:poisSimFig} depicts epidemics on a network with a Poisson degree distribution ($z=1.5$) and parameters $r=0.2,\mu=0.1$, and $\rho=0.25$. Figure~\ref{fig:plSimFig} shows epidemics on a network with a power law degree distribution ($\alpha=2.1, \kappa=75$) and parameters $r=0.2,\mu=0.1$, and $\rho=0.20$.  
		The deterministic NE model (table~\ref{tab:netsir2}) predicts a trajectory which passes through the central-most region of the swarm of simulation trajectories and shows good agreement with the final size. There is nevertheless a great deal of variability among the simulation trajectories in terms of the onset of the expansion phase--the point in time when the epidemic increases at its maximal rate. At the onset of expansion phase, all trajectories are more or less similar, in agreement with the NE model.
		
		In contrast to the homogeneous Poisson network, the power law gives an almost immediate expansion phase. This can be understood by noting that the hazard of infection is proportional to $p_I$, and initially
		\begin{equation}
			\label{eqn:pidotinitial}
			{\displaystyle \dot{p}_I(t=0) = \epsilon\left(r\frac{g''(1-\epsilon)}{g'(1-\epsilon)} -r-\mu-\frac{\rho\epsilon}{1-\epsilon} \right) }
		\end{equation}
		There is a ratio of PGF's in equation~\ref{eqn:pidotinitial}:
		\begin{eqnarray*}
		{\displaystyle g''(1-\epsilon)/g'(1-\epsilon)\approx  g''(1)/g'(1) = \frac{\sum_k k^2 p_k}{\sum_k kp_k} - 1}
		\end{eqnarray*}
		This is approximately the ratio of the second moment to the first moment of the degree distribution, which for the power-law approaches infinity as the cutoff $\kappa\rightarrow \infty$. Because the ratio is very large, power-law networks have almost immediate expansion phase.~\cite{barthBarrSatoVesp1,satoVesp2,satoVesp3,boguSatoVesp1}
		
\section{Application of the NE model to syphilis outbreak among Atlanta adolescents\label{sec:syph}}
We demonstrate the utility of the NE model using link-tracing data from a 1996 outbreak of syphilis among Atlanta adolescents. Sexual network data typically report contact time, duration of a contact, and frequency of interaction. Prior network models have typically not taken into account the serial aspect of sexual contacts, and instead assume that all contacts reported in a survey are constant over the duration of an epidemic or infectious period. Here, we illustrate that the dynamic SIR network model (table~\ref{tab:netsir2}) can explicitly capture the transitory nature of sexual contact patterns. 

We use public-health data from an outbreak of syphilis within an adolescent community centered on an Atlanta high-school~\cite{rothenberg1998usn}. Initially, several adolescents diagnosed with syphilis were interviewed by epidemiologists. The sexual contacts of these respondents were then traced and interviewed. In all, 34 people were interviewed and 204 contacts were traced. Each interviewee named their sexual contacts and listed the date of their first and last interaction with each contact. 

The complexity of syphilis transmission dynamics~\cite{framework1997nhs} and the small size of our data set make modeling the 1996 outbreak quite difficult. The following results should thus be taken with the caveat that there is significant uncertainty in the estimated rates and parameters, particularly the host transmission and recovery rates. Below we will show the impact of these parameters on the expected final size of a syphilis epidemic. 

We estimate the relevant parameters using equations given in the supplement~\cite{supp} and in~\cite{doug,dougSalg1,volz2}. In brief, a typical syphilis infection can last about a year if left untreated, and a typical infectious period will last 154 days on average~\cite{jonesSyphilis}. A convenient estimate of the recovery rate is then $\hat{\mu}=1/154=0.0065$. This estimate ignores many features of the pathology of syphilis for mathematical convenience, such as different probabilities of recovery at different stages of the infectious period~\cite{framework1997nhs}. 
There are diverse estimates for the transmissibility of syphilis, ranging from 9.2\% to 63\% per partner~\cite{framework1997nhs}. The estimate of 62.7\% was selected as the most credible by the authors in~\cite{framework1997nhs}.

Using the contact durations reported in the Atlanta study, we estimate the mixing rate of the population to be $\hat{\rho}=0.032$~\cite{supp}. We then use the reported numbers of contacts to estimate $\hat{k}_{ego}$, the average number of concurrent contacts for each individual $ego$ in the sample.
The degree distribution $p_k$ can then be estimated~\cite{supp,doug,dougSalg1,volz2} from the sequence $\hat{k}_i$, which is well fit by a power law with exponent $\alpha=-2.66$. For the power-law fit, $\chi^2/n = 0.018$. An exponential distribution provided a worse fit to the data with $\chi^2/n=0.632$. We assume the estimated degree distribution in the following analysis.

\begin{figure}
	\includegraphics[width=.5\textwidth]{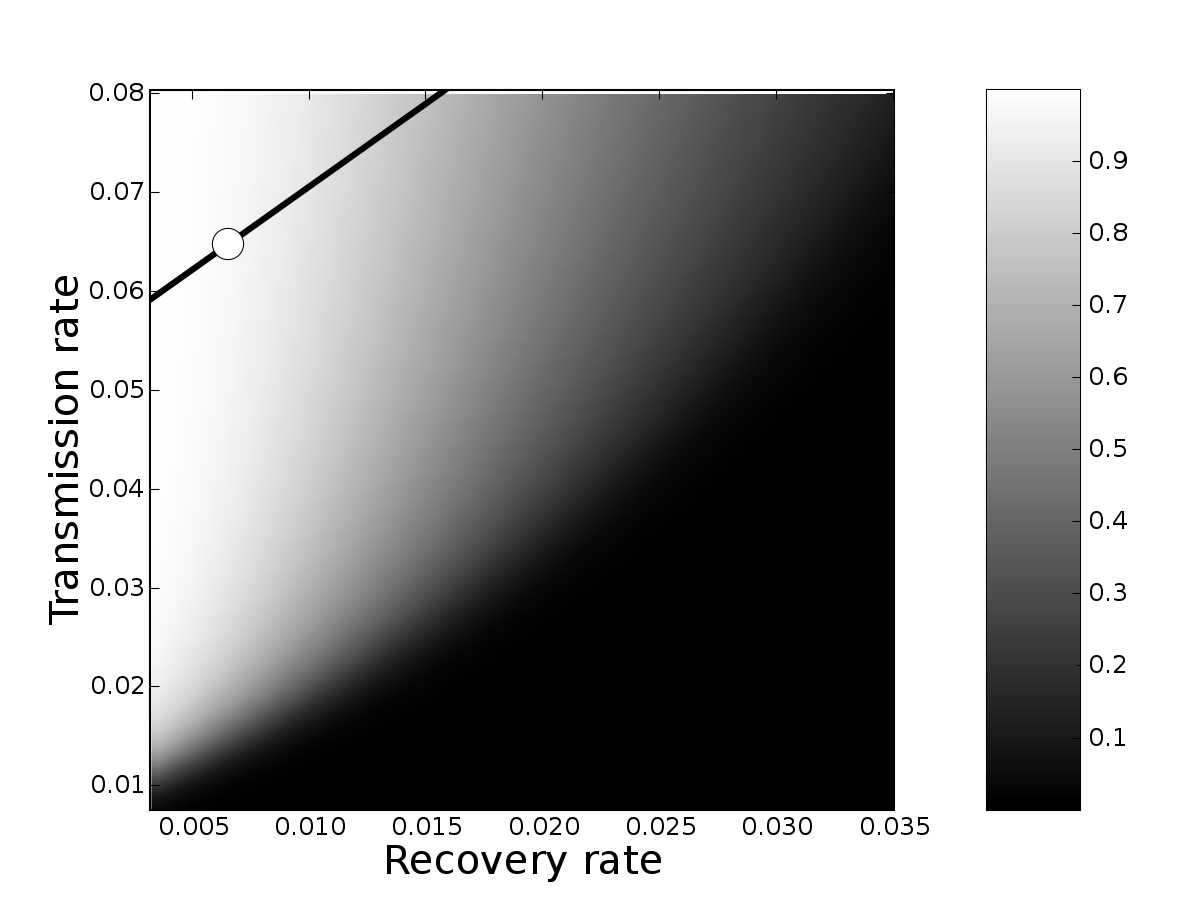} 
	\caption{The final epidemic size as predicted by the NE model (table~\ref{tab:netsir2}) is shown with respect to the transmission rate $r$ and recovery rate $\mu$ for the Atlanta syphilis data. Lighter colors correspond to larger final size, as given by the color bar on right. The thick black line corresponds to the ratio $r/\mu$ that gives the expected transmissibility of $\tau=0.627$. The large data point indicates the expected recovery rate of $\mu=1/154$. \label{fig:fshm}}
\end{figure}

Figure~\ref{fig:fshm} shows the final size of an outbreak as predicted by the NE model over a broad range transmission and recovery rates.  The black line (upper left) corresponds to the ratio of $r$ to $\mu$ that yields a transmissibility $\tau=0.627$. If we suppose a transmissibility of 62.7\% and a recovery rate of $1/154 = 0.0065$ (the white circle in figure~\ref{fig:fshm}), the final size is expected to be 97\%. This is not consistent with the observed outbreak or an estimated prevalence of 35\%~\cite{supp,doug,dougSalg1,volz2}. As figure~\ref{fig:fshm} shows, however, slight alterations in $r$ or $\mu$ can drastically impact the predicted outbreak size. In particular, therapeutic intervention, which certainly occurred during the 1996 outbreak, increases the effective recovery rate and thereby decreases the ultimate attack rate. Therefore the estimate of 97\% should be considered a worst-case scenario given the behavioral parameters (mixing rate and degree distribution) estimated for this adolescent population.

\section{Discussion}
Human contact patterns are characterized by heterogeneous numbers of transitory contacts. If contacts change at a rate which is slow relative the rate of epidemic propagation, then static network approximations such as those based on bond-percolation~\cite{callaway2000nra,newm1,meyers2005nta,meyers2006ped} 
may be appropriate. On the other hand, if contacts have very short duration relative to epidemic dynamics, then static network approximations break down and a mass-action model is more appropriate (equations~\ref{eqn:thetadotrm} and~\ref{eqn:midotrm}). In between these extremes, contacts are neither fixed nor instantaneous, and accurate epidemiological forecasting requires models that explicitly capture their dynamics, such as the NE model developed here.  In fact, by changing a single parameter (the mixing rate), the NE model crosses the spectrum of models from static network to mass-action. 

The NE model is particularly useful for building models from link-tracing data. For many data sets, standard mass action models do not adequately capture the finite number and extended-duration of contacts, while static network models ignore the transitory and serial nature of contacts. Using the example of a 1996 syphilis outbreak in an adolescent population, we showed that the NE model can be easily fit to sexual contact data and then used to explore the epidemiological implications of host population structure.

In the limit of large mixing rate, the NE model becomes a simple (low-dimensional) mass-action model (equations~\ref{eqn:thetadotrm} and~\ref{eqn:midotrm}) that captures SIR dynamics in populations with arbitrary heterogeneity of contact rates. It reduces to the standard mass-action model when one assumes that all individuals have the same number of contacts. The mass-action model could potentially find wide utility in populations which are heterogeneous with respect to contact rates, infectiousness or susceptibility, specifically for modeling highly contagious diseases (such that brief contacts lead to transmission) or slow-propagating infectious diseases (such as many STD's) -- In either case, epidemic propagation is slow relative to the turnover in contacts. 

We check our mathematical results using simulations which model continuous-time stochastic processes (both social and epidemiological) and take into account the finite size and heterogeneity of the population (section ~\ref{sec:sim}). We wish to highlight our specific simulation techniques as an interesting alternative to the commonly used chain-binomial simulation~\cite{daley2001emi}. 

The NE model offers a flexible starting point for analyzing epidemiological processes in dynamic networks. It should be fairly straightforward to extend the model to populations with simple spatial heterogeneity or assortative mixing by type~\cite{newman2003mpn}. Modeling a dynamic population in which the number of concurrent contacts varies in time, however, will likely require a different approach.\\

\emph{The authors thank Richard Rothenberg for providing the Atlanta dataset and useful comments. LAM acknowledges grant support from the James S. McDonnell Foundation.}


\end{document}